\documentclass{emulateapj}
\usepackage{graphicx}
\usepackage{natbib}
\usepackage{subfigure}
\usepackage{amsmath}
\usepackage{url}
\usepackage{bm}
\providecommand{\e}[1]{\ensuremath{\times 10^{#1}}}
\renewcommand{\vec}[1]{\bm{#1}}
\DeclareMathOperator\erf{erf}
\begin{document}

\title{The Close Companion Mass-Ratio Distribution of Intermediate-Mass Stars}

\author{Kevin Gullikson \altaffilmark{1}}
\author{Adam Kraus \altaffilmark{1}}
\author{Sarah Dodson-Robinson \altaffilmark{2}}

\altaffiltext{1}{University of Texas, Astronomy Department. 2515 Speedway, Stop C1400. Austin, TX 78712}
\altaffiltext{2}{University of Delaware, Department of Physics and Astronomy, 217 Sharp Lab, Newark, DE 19716}

\begin{abstract}
Binary stars and higher-order multiple systems are an ubiquitous outcome of star formation, especially as the system mass increases. The companion mass-ratio distribution is a unique probe into the conditions of the collapsing cloud core and circumstellar disk(s) of the binary fragments. Inside $a \sim 1000$ AU the disks from the two forming stars can interact, and additionally companions can form directly through disk fragmentation. We should therefore expect the mass-ratio distribution of close companions ($a \lesssim 100$ AU) to differ from that of wide companions. This prediction is difficult to test using traditional methods, especially with intermediate-mass primary stars, for a variety of observational reasons. We present the results of a survey searching for companions to A- and B-type stars using the direct spectral detection method, which is sensitive to late-type companions within $\sim 1''$ of the primary and which has no inner working angle. We estimate the temperatures and surface gravity of most of the 341 sample stars, and derive their masses and ages. We additionally estimate the temperatures and masses of the 64 companions we find, 23 of which are new detections. We find that the mass-ratio distribution for our sample has a maximum near $q \sim 0.3$. Our mass-ratio distribution has a very different form than in previous work, where it is usually well-described by a power law, and indicates that close companions to intermediate-mass stars experience significantly different accretion histories or formation mechanisms than wide companions.
\end{abstract}

\maketitle

\section{Introduction}

\label{sec:intro}

Stellar multiplicity is an inevitable and common outcome of star formation, with roughly half of all solar-type field stars in binary or multiple systems \citep{Raghavan2010} and an even higher fraction as the stellar mass increases \citep{Zinnecker2007}. Young stellar associations and clusters tend to have even higher multiplicity \citep{Duchene2013}, indicating that stars often form in multiple systems that are subsequently destroyed by dynamical interactions as the cluster dissociates. 

The overall multiplicity rate and the distributions of mass ratio, period, and eccentricity of a binary star population place important constraints on the mode of binary star formation. While the period and eccentricity are altered by dynamical processing in the birth cluster, the present-day mass ratio of a binary system is a direct result of its formation \citep{Parker2013}. Most binary stars are thought to form via core fragmentation \citep{Boss1979, Boss1986, Bate1995}, in which a collapsing core fragments into two or more individual protostars. The number and initial masses of the fragments are set by the total core mass, as well as its rotation, turbulence, and its temperature and density structure. If the fragments are well separated ($a \gtrsim 1000$ AU), they will evolve independently of each other, accreting mass from the core material onto their own protostellar disks and then onto the protostars themselves. However close fragments ($a \sim 100$ AU) will interact with each other; the protostellar disk may be truncated, destabilized, or form into a circumbinary disk if the separation is small enough \citep{Bate1997}. In addition, an unstable disk can fragment to form low-mass companions \citep{Kratter2006, Stamatellos2011}. The mass ratios of close companions formed via either mechanism should be affected by preferential accretion. Most work has suggested that the disk material will preferentially accrete onto the lower mass companion \citep{Bate1997, BBB2002}; however, recent work has indicated that magnetic disk braking may result in preferential accretion onto the more massive component \citep{Zhao2013} instead. In either case, we would expect to find a mass-ratio distribution for companions inside a few $100$ AU that differs from that of companions on wider orbits where preferential accretion does not occur.

The mass ratio, period, and eccentricity distributions are well-known for solar type stars \citep{Duquennoy1991, Raghavan2010} and cooler stars \citep{Fischer1992, Delfosse2004}. Interestingly, \citet{Reggiani2011}, and later \citet{Meyer2013}, found that the mass-ratio distribution of field solar-type and M-dwarf stars is invariant to separation. The field M-dwarf semimajor axis distribution peaks near $\sim 5$ AU \citep{Duchene2013}, with very few companions at separation $\gtrsim 100$ AU; the 27 stars used in the analysis by \citet{Reggiani2011} is unsufficient to compare the mass-ratio distribution inside $\sim 100$ AU with that outside it. However, the solar-type period distribution peaks near $45$ AU \citep{Raghavan2010}, with roughly $40\%$ of binary systems on orbits wider than $100$ AU. The nondetection of a difference in mass-ratio distribution is significant, although with only 30 stars in the field sample it is difficult to completely rule out that such a difference exists.

All of the orbital distributions are much less certain for more massive stars. The reason for this is two-fold: first, more massive stars tend to be more distant than sunlike or low-mass stars, meaning many of the companions are angularly close to the primaries and difficult to detect with imaging techniques. Second, the primary stars tend to be rapid rotators, which limits radial velocity precision to $\sim 1 \mathrm{km\ s}^{-1}$ and causes the spectral lines of double-lined systems to blend. Radial velocity monitoring can only measure a mass ratio if spectral lines from both components are visible and separable; this typically suffers from the same flux ratio difficulty as imaging techniques. 

Nonetheless, \citet{DeRosa2014} performed an adaptive optics imaging survey of nearby A-type stars, and found that the mass-ratio distribution is well-described by a power law with large slope, indicating a very strong preference for low-mass companions. They also found initial evidence that the mass-ratio distribution for companions inside $125$ AU has a much shallower power law slope than that of wide companions, and is consistent with flat. Their close companion subsample contained only 18 binary systems, and the result is complicated by the inherent difficulty of detecting close companions with low mass ratios in an imaging survey. 

Radial velocity monitoring surveys can detect much closer companions than imaging surveys, but are typically only complete to low-mass companions if the primary is a slow rotator. Chemically peculiar Am stars are typically associated with binary companions, and are slow rotators due to tidal braking; they thus form a highly biased sample of intermediate-mass stars. Nonetheless, it is interesting to note that they have a mass-ratio distribution which peaks near $q \sim 0.5$ \citep{Vuissoz2004}, an entirely different form than the distribution found around chemically normal stars at wide separations.

In this paper, we describe a spectroscopic survey of nearby chemically normal, main sequence intermediate-mass stars ($M \approx 1.5 - 15 M_{\odot}$). We search for companions using the direct spectral detection technique \citep{Gullikson2016}, which has a separation-invariant detection rate for all separations inside $\sim 1 ''$. We describe the stellar sample and data used for the survey in Section \ref{sec:obs}, as well as the data reduction steps in the same section. Next, we describe the direct spectral detection method and tabulate the companion detections in Section \ref{sec:companions}. We estimate the mass and age of the sample stars in Section \ref{sec:sp}, and discuss the survey completeness in Section \ref{sec:completeness}. Finally, we end with a derivation of the mass-ratio distribution from our sample in Section \ref{sec:mrd} and discuss its implications for binary formation in Section \ref{sec:discussion}.

\section{Observations and Data Reduction}
\label{sec:obs}

The stellar sample for this survey is defined by the following criteria:

\begin{itemize}
\item $V < 6$ mag
\item $v\sin{i} > 80 \mathrm{km\ s}^{-1}$
\item Spectral Type A or B with the following additional constraints
\begin{itemize}
  \item Main Sequence
  \item No spectral peculiarities except for `n', which denotes broad lines.
\end{itemize}
\end{itemize}

The magnitude limit ensures that a sufficiently high signal-to-noise ratio can be achieved in a short period of time. It does introduce a Malmquist bias in the derived mass ratio, which we discuss and correct for in Section \ref{sec:mrd}. Likewise, the $v\sin{i}$ limit makes accounting for the primary star spectrum in the companion search trivial; since most A- or B-type stars are rapid rotators, the cutoff removes less than half of the stars from the potential sample. We exclude pre-main sequence stars because both the primary and the companion mass would depend very strongly on young and uncertain ($\lesssim 1$ Myr) evolutionary models. Finally, we exclude post-main sequence stars from our sample because the binary flux ratio would be even less favorable to companion detection in an evolved star. Most of the spectral peculiarities denote narrow lines, which are already removed from the sample by the $v\sin{i}$ cut. The sample is given in Table \ref{tab:sample}. The spectral type, coordinates, V magnitude, and parallax are all adopted from the Simbad Database \citep{Simbad}, while the stellar effective temperature, surface gravity, masses, and ages are discussed in Section \ref{sec:sp}. 

The sample, being comprised of early-type stars, is heavily biased towards young stars. The estimated ages range from about 10 Myr to 1 Gyr, with most falling in the range of a few tens or hundreds of Myrs. The sample also mostly contains nearby stars, although the magnitude limit provides a greater extent than a volume-limited survey would. The parallactic distances in our sample range from $\sim 15 - 2000$ pc. The maximum detectable separation, which we define as the point at which the companion is no longer guaranteed to fall on the spectrograph slit (see Section \ref{subsec:specdata} for a description of the spectrographs we use), is $\sim 20 - 4000$ AU. The median parallactic distance in our sample is $95$ pc, corresponding to a projected separation of $\sim 200$ AU. Most of the companions we are able to detect are close enough to have been impacted by the circumprimary disk, with $85\%$ of the sample sensitive to companions inside $100$ AU.

\subsection{Spectroscopic Data}
\label{subsec:specdata}
We use several high spectral resolution, cross-dispersed \'echelle spectrographs for this survey. We use the CHIRON spectrograph \citep{CHIRON} on the 1.5m telescope at Cerro Tololo Inter-American Observatory for most southern targets. This spectrograph is an $R\equiv \lambda / \Delta \lambda = 80000$ spectrograph with wavelength coverage from 450 - 850 nm, and is fed by a $2.7''$ optical fiber. The data are automatically reduced with a standard CHIRON data reduction pipeline, but the pipeline leaves residuals of strong lines in adjacent orders. We therefore bias-correct, flat-field and extract the spectra with the optimum extraction technique \citep{Horne1986} using standard IRAF\footnote{IRAF is distributed by the National Optical Astronomy Observatories, which are operated by the Association of Universities for Research in Astronomy, Inc., under cooperative agreement with the National Science Foundation.} tasks, and use the wavelength calibration from the pipeline reduced spectra.

For the northern targets, we use a combination of the High Resolution Spectrograph \citep[HRS,][]{HRS} on the Hobby Eberly Telescope, and the Tull coud\'e \citep[TS23,][]{TS23} and IGRINS \citep{IGRINS} spectrographs, both on the 2.7m Harlan J. Smith Telescope. All three northern instruments are at McDonald Observatory. For the HRS, we use the $R = 60000$ setting with a $2''$ fiber, and with wavelength coverage from 410-780 nm. We bias-correct, flat-field, and extract the spectra using an IRAF pipeline very similar to the one we use for the CHIRON data. The HRS spectra are wavelength-calibrated using a Th-Ar lamp observed immediately before or after the science observations.

For the TS23 spectrograph, we use a $1.2''$ slit in combination with the E2 \'echelle grating (53 grooves/mm, blaze angle $65^{\circ}$), yielding a resolving power of $R=60000$ and a wavelength coverage from 375-1020 nm. We reduce the data using an IRAF pipeline very similar to the ones we use for CHIRON and HRS, and wavelength calibrate using a Th-Ar lamp observed immediately before the science observations.

IGRINS has a single setting with $R = 40000$. It has complete wavelength coverage from $1475-2480$ nm, except in the telluric water band from $1810 - 1930$ nm. Each star is observed in an ABBA nodding mode, and reduced using the standard IGRINS pipeline \citep{IGRINS_plp_v2}. The standard pipeline uses atmospheric OH emission lines as well as a Th-Ar calibration frame to calibrate the wavelengths; we further refine the wavelength solution using telluric absorption lines in the science spectrum.

After reducing the data, we fit and remove the telluric spectrum using the TelFit code \citep{Gullikson2014}. We fit each \'echelle order affected by telluric absorption independently from each other to get the best removal. The telluric correction is critical for IGRINS spectra, where every order is dominated by telluric absorption lines. For the optical spectra, it is less critical but allows us to use some of the redder orders than we otherwise would be able to. For unsaturated lines, the best-fit telluric model reproduces the data to within $\sim 1-5\%$ of the continuum level.

We give the spectroscopic observation log in Table \ref{tab:observations}. We calculate the signal-to-noise ratio (the ``snr'' column) for the optical instruments (CHIRON, TS23, and HRS) as the median of the extracted flux divided by its uncertainty for each pixel from the \'echelle order nearest $675$ nm. For the IGRINS instrument, we calculate the signal-to-noise ratio from the order nearest $2200$ nm.

\subsection{Imaging Data}
As part of the follow-up effort, we used the NIRI instrument behind the Altair adaptive optics system on the Gemini North Telescope. For each star listed in Table \ref{tab:imaging_obs}, we obtained 25 images in 5 dithering positions. We used the K-continuum band centered on $2.2718\ \mu m$ and a variety of exposure times and dates (listed in Table \ref{tab:imaging_obs}). Because the targets are all extremely bright, we used the high read noise and high flux detector settings to allow for very short co-add exposure times and to prevent saturation. We reduced the data using the Gemini set of IRAF tasks, which include steps for nonlinearity correction, flat-fielding, sky subtraction, and co-addition of the dither frames. 

We measure the flux and position of both stars by fitting a 2D Moffat function \citep{Moffat1969} to both stars simultaneously, constraining the shape parameters for both functions to be the same. The ratio of the amplitudes gives the magnitude difference, and the pixel locations along with the detector pixel scale gives the separation and position angle between the stars. We note that the goal of these images was confirmation and we did not observe any reference targets to make a distortion map and correct the image rotation. The uncertainty in position angle and to a lesser degree separation quoted in Table \ref{tab:imaging_obs} is likely underestimated.

\begin{figure*}
\includegraphics[width=\textwidth]{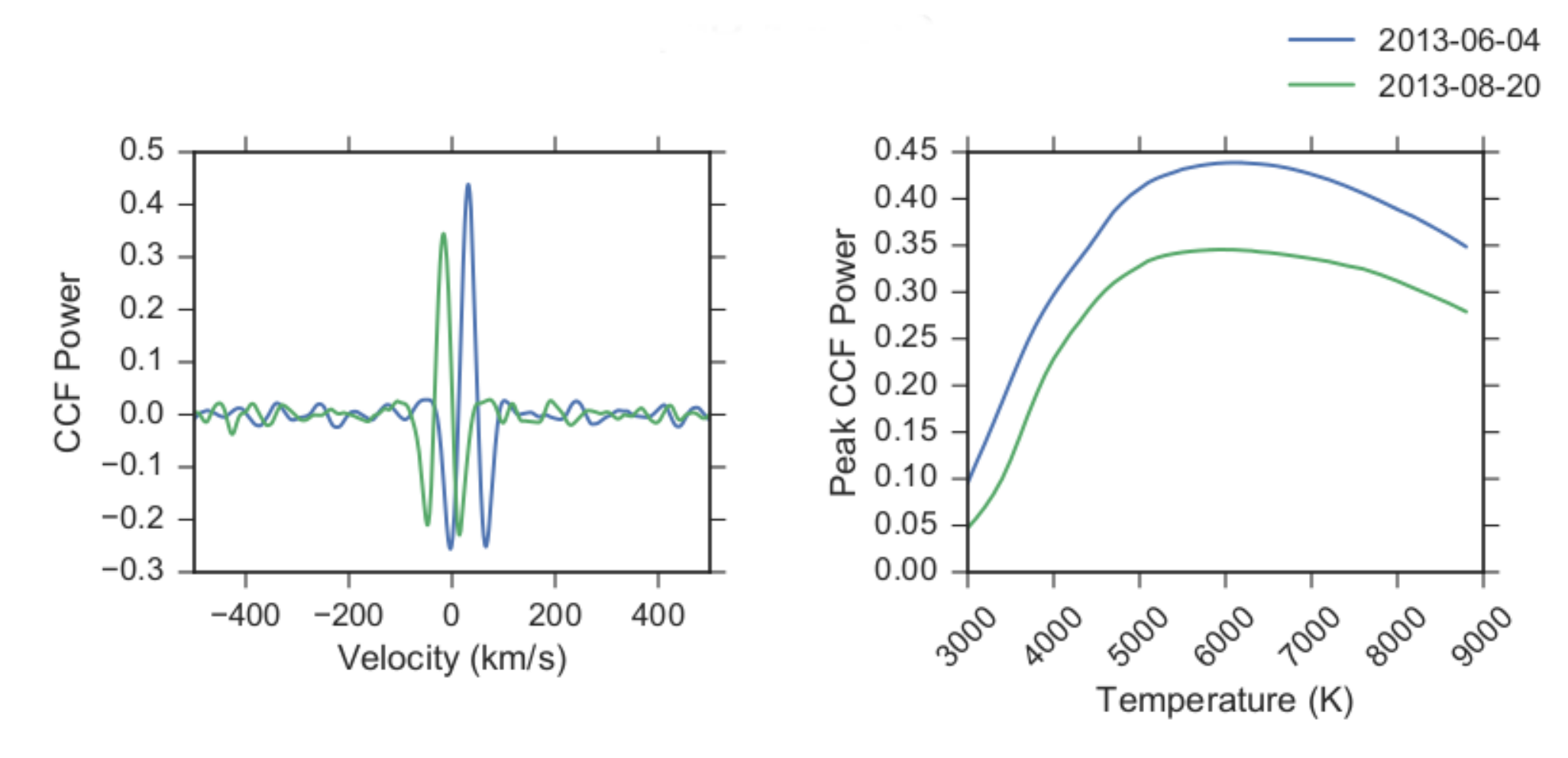}
\caption{\emph{Left}: Cross-correlation function between the observed spectra of HIP 109139 and a $5700\ K$ Phoenix model spectra. The detection at two dates shows significant velocity variation, indicating orbital motion with a short period. \emph{Right}: Peak CCF height as a function of Phoenix model spectra template temperature. The maxima of the curves indicate the temperature of the companion.}
\label{fig:ccf}
\end{figure*}

\section{Companion Search}
\label{sec:companions}

We search for stellar companions to our sample stars using the direct spectral detection technique, described in detail in \citet{Gullikson2016}. In short, we unsharp-mask each spectrum using a gaussian filter with width proportional to the primary star $v\sin{i}$ to remove the broad lines from the primary star. We then cross-correlate each \'echelle order of each filtered spectrum against a large grid of Phoenix model spectra \citep{Husser2013_b} with the following parameters:

\begin{itemize}
\item $T_\mathrm{eff} = 3000-12000$ K, in steps of 100 K
\item {[}Fe/H{]} = -0.5, 0.0, +0.5
\item $v\sin{i} = 1, 5, 10, 20, 30 \ \mathrm{km\ s}^{-1}$
\end{itemize}

In order to be sensitive to hot companions, we additionally cross-correlate the spectra against a second grid of Kurucz model spectra \citep{Castelli2003}. The change in model is necessary because the Phoenix model library does not extend beyond $12000\ K$. The Kurucz grid is defined as follows:

\begin{itemize}
\item $T_\mathrm{eff} = 9000-30000$ K, in steps of 1000 K
\item {[}Fe/H{]} = -0.5, 0.0, +0.5
\item $v\sin{i} = 1, 5, 10, 20, 30, 40, 50 \ \mathrm{km\ s}^{-1}$
\end{itemize}

We combine the cross-correlation functions for all orders using both a simple average and the maximum-likelihood weighting scheme \citep{Zucker2003}. A companion detection is denoted by a strong peak in the combined cross-correlation function (CCF). While the maximum-likelihood scheme produces detections with much higher significance, it also magnifies spurious peaks and so has a larger false-positive rate. For this reason, we use the simple average CCFs in all further analysis.

The peak height in the CCF as a function of the stellar model acts in a similar way to the more typical $\chi^2$ map of parameter space. More concretely, as the stellar model template gets closer to the true companion spectrum, the CCF peak gets higher. We can therefore measure the companion temperature and, to a lesser degree its metallicity and $v\sin{i}$, in a single spectrum. We calculate the measured temperature ($T_m$) and variance ($\sigma_T^2$) as a weighted sum near the grid point with the highest CCF peak value, weighting by the peak CCF height at each temperature ($C_i$):

\begin{eqnarray}
\label{eqn:tmeas} 
T_m &=& \sum_i C_i T_i / \sum_i C_i \\
\sigma_T^2 &=& \frac{\sum_i C_i (T_i - T_m)^2}{ \sum_i C_i - \sum_i C_i^2 / \sum_i C_i}
\end{eqnarray}
Typical uncertainties are on the order of $200$ K. In the case of multiple observations for the same star, we use the variance-weighted mean of the individually measured temperatures.

Imperfect stellar models cause a bias between the true companion temperature and the measured temperature ($T_m$). This bias is most pronounced at low temperatures, where the difficult-to-model molecular absorption becomes important. We correct for the bias by applying the linear calibrations developed in \citet{Gullikson2016}. These calibrations are only valid for companions with $3000 < T_\mathrm{eff} < 7000 K$; for detections at hotter temperatures we assume that the temperature which produces the maximum CCF peak is an \emph{unbiased} estimator of the true companion temperature.

We list the companion detections in Table \ref{tab:companions}, and report the estimate of the companion temperature, $v\sin{i}$, and metallicity derived from the model parameters which produce the largest CCF peak. The $v\sin{i}$ and metallicity values do not have uncertainties and should only be taken as a rough estimate of the true value. The mean and standard deviation of the companion metallicities is $-0.29 \pm 0.30$; the marginal bias towards low metallicities is most likely a measurement bias and does not reflect the true companion population \citep{Gullikson2016}.  We show the detection CCFs and a plot of peak CCF height as a function of model temperature for HIP 109139 in Figure \ref{fig:ccf}. Similar figures for all companions are available in the supplementary files.

We have follow-up spectroscopy for 15/23 of the new companions to confirm their existence. In most cases, there is a clear shift in the radial velocity of the companion, indicating that it is orbiting the target star and is not a foreground or background contaminant (See Figure \ref{fig:ccf}). Two of the new detections (companions to HIPs 38593 and 79404) were detected twice but not in a third attempt, most likely because the third spectrum had low signal-to-noise. The companion to HIP 93805, at $\sim 4000$ K, was detected twice with near-infrared IGRINS but not the optical CHIRON instrument that is less sensitive to cool companions than IGRINS. Two of the companions with only one detection were observed at least twice (HIPs 19949 and HIP 23362); both of the non-detections are from the IGRINS instrument, which is less sensitive to hot companions with rapid rotation speeds because there are far fewer spectral lines of the companion in the near-infrared than there are in the optical.

\begin{figure}
\includegraphics[width=\columnwidth]{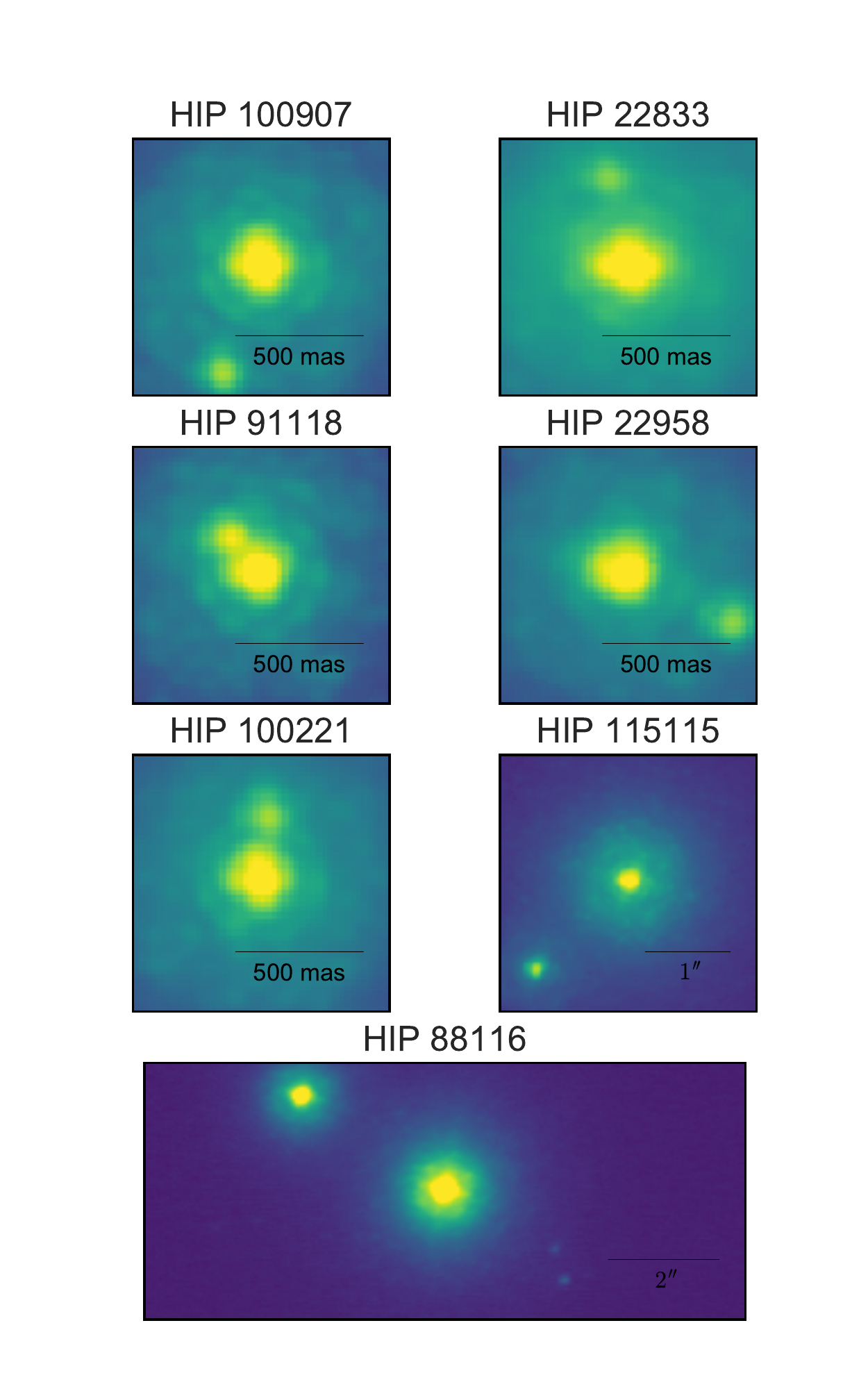}
\caption{Detection images for all stars in which we detect a companion in the follow-up NIRI data. 
There are several nearby sources for HIP 88116, none of which are the source we detect in the 
spectroscopic data (see Section \ref{sec:companions}).}
\label{fig:images}
\end{figure}

In addition to the spectroscopic follow-up, we obtained Gemini/NIRI adaptive optics imaging data for 18 of the northern companions, and were able to resolve 7 of them. We show the separation, position angle, and magnitude difference measurements in Table \ref{tab:imaging_obs}, and display the images in Figure \ref{fig:images}. We also derive the projected separation in AU and the companion mass from the images. We calculate the separation from the measured angular separation and the Hipparchos parallax \citep{VanLeeuwen2007}. We calculate the companion mass and uncertainties from 1000 samples of the magnitude difference measurement and the primary star mass, temperature, age, and radius (see Section \ref{sec:sp}). For each sample, we use a grid of Kurucz stellar model spectra \citep{Castelli2003}and the pysynphot code\footnote{pysynphot is a python package to perform synthetic photometry, and is available at this url: \url{https://pypi.python.org/pypi/pysynphot}} to determine the companion temperature needed to replicate the observed magnitude difference. We estimate the companion radius by interpolating solar metallicity Dartmouth isochrones \citep{Dotter2008} from the companion temperature and system age sample. We convert the best temperature to a companion mass using the same isochrone grid. The masses derived from the imaging data have very large uncertainties because the primary star property estimates that they depend on are very uncertain. The imaging masses agree with the spectroscopically-derived masses in Table \ref{tab:companions}, with the exception of HIP 115115 which has a much higher mass from the imaging data than the spectroscopic data. The spectroscopic masses for all stars are more reliable, since they are less model-dependent.

One star in the imaging sample, HIP 88116, has several nearby sources in the image. We quote the magnitude difference and separation of the brightest source in Table \ref{tab:imaging_obs}, but stress that \emph{none} of the visible sources is likely to be the companion we see in the spectroscopic data. The two epochs of spectroscopic data show a radial velocity shift of $\Delta v = 30.7\ \mathrm{km\ s}^{-1}$ over the course of roughly one year; this orbital motion is much too large to be consistent with any of the companions visible in the image (all with separations $ > 1''$ and projected separations $ > 300$ AU).

\section{Sample Star Parameters}
\label{sec:sp}

In order to convert from companion temperature to mass ratio, we first need an estimate of the primary mass. In addition, since the primary stars in our survey have short main-sequence lifetimes, some companions may still be contracting onto the main sequence and so an age estimate for the system is necessary to convert from companion temperature to mass.

About half of our sample stars have robust mass and age estimates from Str\"omgren uvby$\beta$ photometry \citep{David2015}. For those that do not, we estimate the mass and age of the system from our spectra. We first cross-correlate the data against a grid of solar metallicity Kurucz model spectra \citep{Castelli2003} spanning

\begin{itemize}
\item $7000\ K < T_\mathrm{eff} < 30000\ K$ in steps of $500\ K$ for $T < 10000\ K$, and in steps of $1000\ K$ for hotter templates.
\item $3.0 < \log{g} < 4.5$ in steps of $0.5$ dex
\item $75 < v\sin{i} < 300\ \mathrm{km s}^{-1}$ in steps of $25\ \mathrm{km s}^{-1}$
\end{itemize}

For the optical data, we use the blue \'echelle orders ($\lambda < 5550 \AA$). We ignore the strong hydrogren Balmer lines in the spectrum because they span several \'echelle orders and make continuum normalization very difficult, potentially biasing the result. There are sufficient metal lines in the optical spectra that the resulting CCF always has a very strong peak at the radial velocity of the primary star. The near-infrared IGRINS spectra have very few strong metal lines; we use the subset from $1.51 - 1.73 \micron$ that is dominated by hydrogren Brackett lines for these spectra. Similar to the companion search, we estimate the temperature and surface gravity of the stars from the CCF with the largest peak. We adopt the following errors on the temperature and surface gravity, which are based on the grid step size and are somewhat more pessimistic than typical uncertainties seen in the literature for A- and B-type stars \citep[e.g.][]{Aydi2014, David2015}:

\begin{align}
 \sigma_T &= \begin{cases}
      \hfill 500\ K \hfill & T < 10000\ K \\
      \hfill 1000\ K \hfill & T >= 10000\ K \\
     \end{cases} \\
 \sigma_{\log{g}} &= 0.25
\end{align}
The IGRINS parameters are less reliable because they rely almost solely on the hydrogren Brackett lines that span an entire \'echelle order, so we double the uncertainty on the IGRINS-derived temperature and surface gravity. Additionally, we throw out the IGRINS parameters if the star was also observed by one of the optical instruments in our survey. For stars observed multiple times, we use the average parameters and reduce the uncertainties accordingly.

\begin{figure}
\includegraphics[width=\columnwidth]{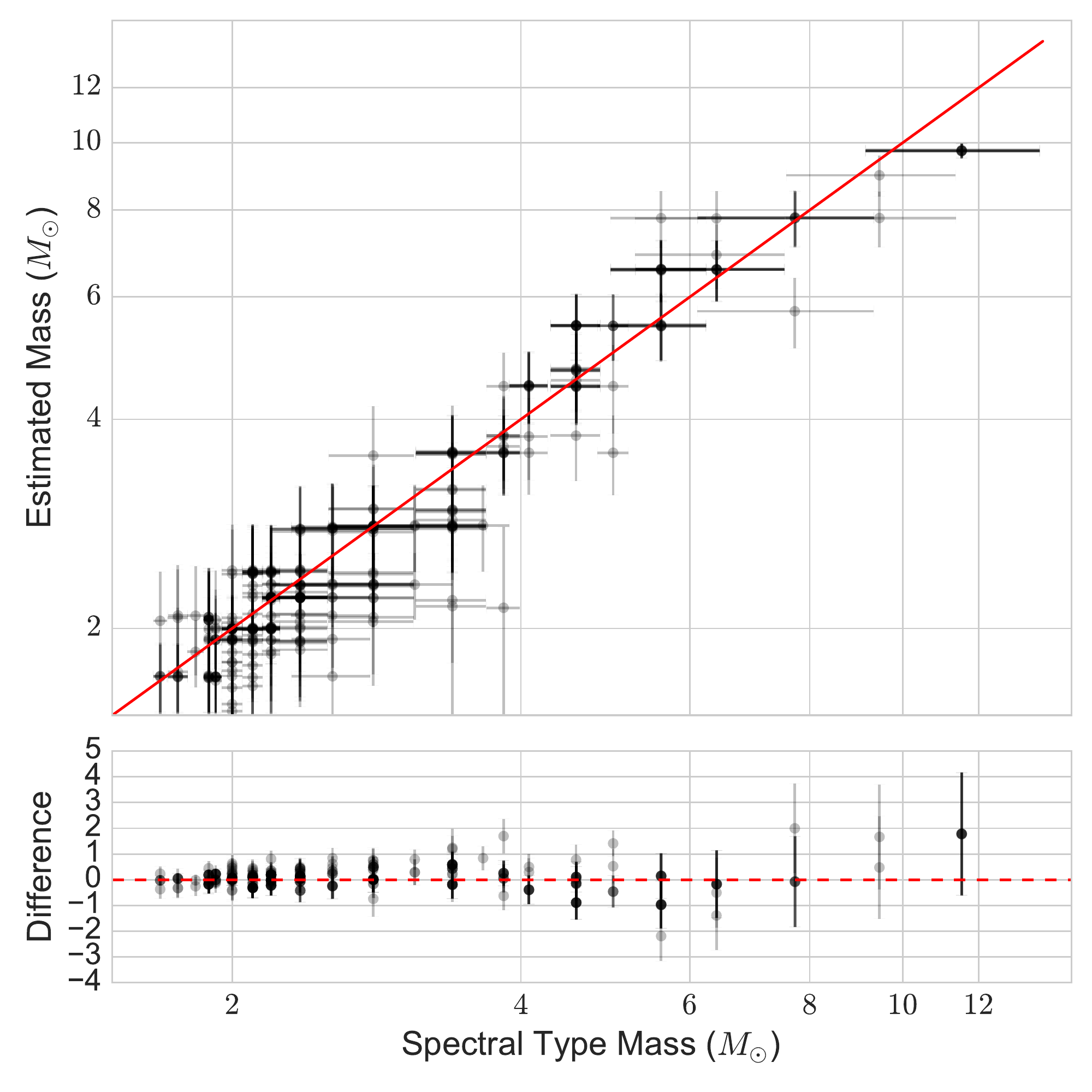}
\caption{Comparison of primary star masses derived from our cross-correlation analysis and Padova isochrones \citep{Bressan2012} with those expected from the published spectral type. There is excellent agreement between the two measures across the entire range of masses.}
\label{fig:prim_mass}.
\end{figure}

Next, we use Padova stellar evolutionary tracks \citep{Bressan2012} and the isochrones code \citep{isochrones_code} to estimate the mass and age of the system from the measured temperature and surface gravity. As a consistency check, we also interpolate from a table of stellar properties as a function of spectral type \citep{Pecaut2013} to estimate the primary mass from the published spectral types. We show the comparison in Figure \ref{fig:prim_mass}. We estimate uncertainties in the spectral type mass by assuming a spectral type uncertainty of $\pm 0.5$ spectral types and propagating to mass. There is excellent agreement between the masses we measure and the spectral type masses.

We show the temperature, surface gravity, mass, and age estimates for most of our sample stars in Table \ref{tab:sample}. We do not give parameters for the few stars that show strong discrepancies with the spectral-type estimate, most of which are early B-stars that have temperatures higher than the maximum grid temperature of $30000\ K$.

\section{Survey Completeness}
\label{sec:completeness}

The detectability of a companion mostly depends on its temperature: cooler companions emit much less light and so are increasingly lost in the Poisson noise from the primary star spectrum. A companion with a high rotation rate is also more difficult to detect because the cross-correlation function gets most of its power from narrow spectral lines.  

\subsection{Injection and Recovery Tests}

To quantify the detection rate as a function of companion temperature and $v\sin{i}$, we performed a series of injection and recovery experiments. We started by creating synthetic binary star observations from each of our observed spectra. We made two distinct grids of companion stars: a low temperature grid spanning

\begin{itemize}
\item $3000\ K < T_\mathrm{eff} < 6500\ K$ in steps of $100\ K$
\item $0\ \mathrm{km\ s}^{-1} < v\sin{i} < 50\ \mathrm{km\ s}^{-1}$ in steps of $10\ \mathrm{km\ s}^{-1}$
\end{itemize}
and a high temperature grid spanning

\begin{itemize}
\item $7000\ K < T_\mathrm{eff} < 12000\ K$ in steps of $1000\ K$
\item $100\ \mathrm{km\ s}^{-1} < v\sin{i} < 250\ \mathrm{km\ s}^{-1}$ in steps of $50\ \mathrm{km\ s}^{-1}$
\end{itemize}
For each grid point, we added a solar metallicity Phoenix model spectrum to the observed data after scaling to replicate the expected flux between a main sequence companion of the model temperature and the known target star spectral type. If the target star had known companions within $3''$, we included the expected flux from the companion when computing the flux ratio. We repeated each grid point at different radial velocities spanning $-400\ \mathrm{km\ s}^{-1} < v < 400\ \mathrm{km\ s}^{-1}$ in $50\ \mathrm{km\ s}^{-1}$ steps to sample the noise properties of the spectra and estimate a probability of detection at each point.

Next, we cross correlated all of the synthetic observations against the Phoenix model template that was used to construct them. We counted the companion as detected if the highest point in the resulting CCF was found at the correct radial velocity, and if the peak had a significance of $>5\sigma$, where $\sigma$ is the standard deviation of the CCF for points more than $100\ \mathrm{km\ s}^{-1}$ away from the peak. We combined all of the radial velocity trials for each grid point to estimate a probability of detection at that grid point:

\begin{equation}
P(\mathrm{detection}) = \frac{N_\mathrm{detected}}{N_\mathrm{rv}}
\end{equation}
where $N_\mathrm{rv} = 17$ is the number of radial velocity trial points. 

\begin{figure}
\includegraphics[width=\columnwidth]{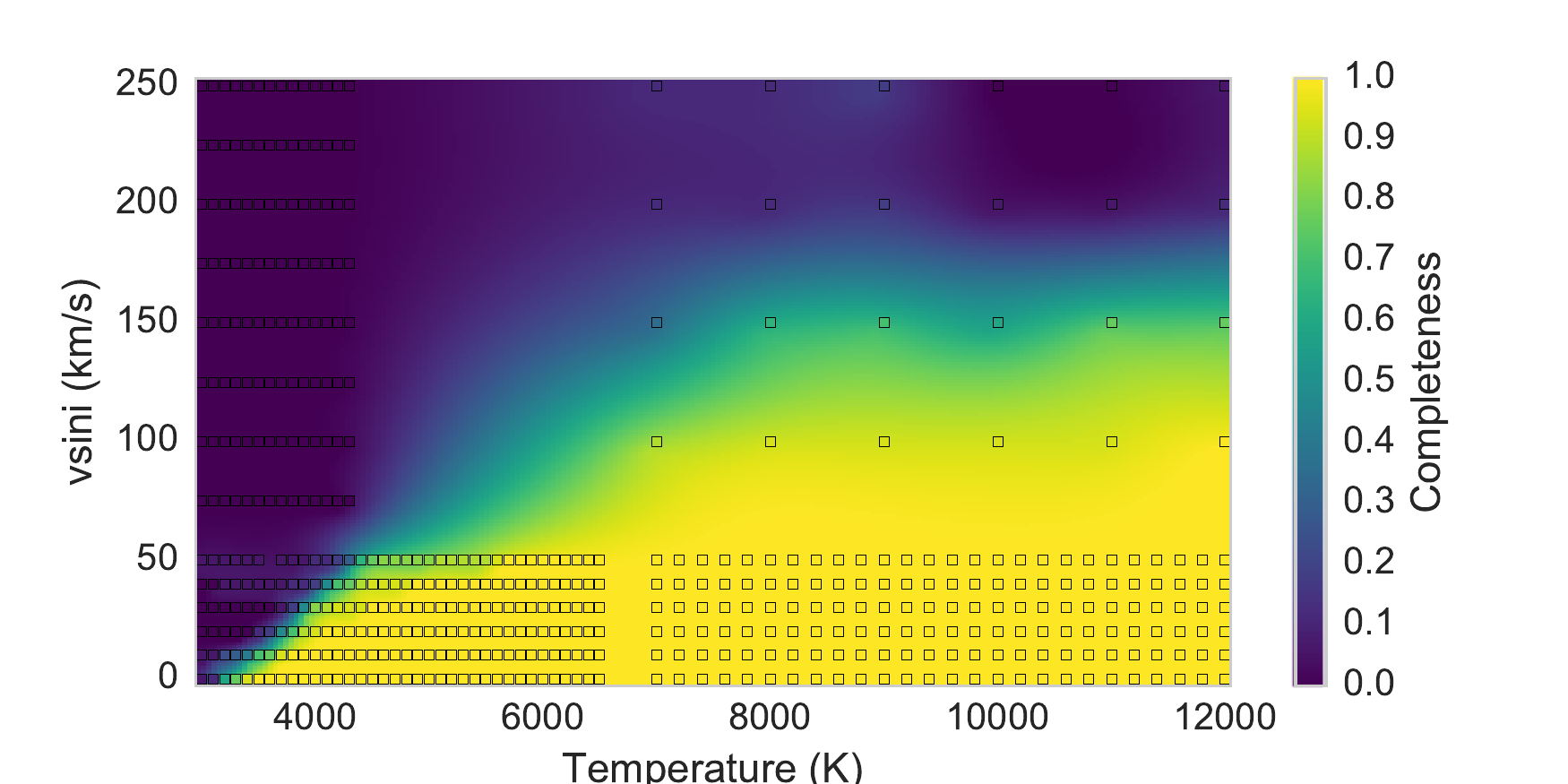}
\caption{Detection rate as a function of companion temperature and $v\sin{i}$ for HIP 24244. All companions that are shaded yellow are detectable, while companions in the purple region are never detectable. The grids of squares in the lower left and upper right show the low temperature and high temperature grid points we used in the sensitivity analysis. The remaining squares come from assumptions about the shape of the detection rate surface and allow us to fully interpolate (see text for details).}
\label{fig:detrate_2d}
\end{figure}

Finally, we interpolated between the grid points using a linear radial basis function interpolator (Figure \ref{fig:detrate_2d}). In order to extrapolate from our grids to estimate the detection rate at high temperature and low $v\sin{i}$ and at low temperature and high $v\sin{i}$, we made the following assumptions about the shape of the two-dimensional detection rate surface: First, we assume that if no companions are detected at temperature $T$ and rotation speed $v\sin{i} = 50\ \mathrm{km\ s}^{-1}$, then no companions will be detected at the same temperature and faster rotation speeds (upper left points in Figure \ref{fig:detrate_2d}). Likewise, we assumed that if all companions are detected at temperature $T=6500$ K and rotation speed $v\sin{i}$, then all companions with the same $v\sin{i}$ and larger temperature will also be detected (lower right points in Figure \ref{fig:detrate_2d}). The detection rate of hot companions is affected by two factors: the increased flux from the companion makes detection easier, while the decreasing number of spectral lines in the companion spectrum makes detection more difficult. If the latter factor is more important, then the assumption we made about the shape of the detection surface is incorrect. We therefore tested the assumption with a small subset of injection and recovery tests, and found that the second assumption we make is valid.

Figure \ref{fig:detrate_2d} shows a clear diagonal dividing line between hot, slow rotators that are always detected and cool, fast rotators that never are. Additionally the figure shows that very fast rotators are never detected, regardless of their temperature, because the signal is completely removed when we unsharp-mask the data (see Section \ref{sec:companions}).

\subsection{Marginalization}
By sampling a suitable distribution of $v\sin{i}$ values for a star of each temperature, we marginalize over the rotation speed:

\begin{equation}
Q(T) = \sum_k Q(T, v_k) v_k 
\end{equation} 
where $Q(T, v)$ is the surface plotted in Figure \ref{fig:detrate_2d} and $v_k$ are the samples from the distribution of $v\sin{i}$. For $T < 6000$ K, we sample $v\sin{i}$ using the gyrochronology relation given in \citet{Barnes2010b}:

\begin{equation}
\frac{k_Ct}{\tau} = \ln\left ( \frac{P}{P_0} \right ) \frac{k_Ik_C}{2\tau^2} (P^2 - P_0^2)
\label{eqn:gyro}
\end{equation}
In Equation \ref{eqn:gyro}, $k_C$ and $k_I$ are constants fit to data with known ages and rotation periods, $P$ and $P_0$ are respectively the current and zero-age main sequence (ZAMS) rotation periods, $\tau$ is the convective turnover time scale and $t$ is the current age of the star. We use the same values that \cite{Barnes2010b} use for the constants:

\begin{itemize}
\item $k_C = 0.646$ day/Myr
\item $k_I = 452$ Myr/day
\end{itemize}
We estimate the convective timescale ($\tau$) by interpolating Table 1 of \citet{Barnes2010a}. We then randomly draw a system age $t$ from its probability distribution function (see Section \ref{sec:sp} and Table \ref{tab:sample}). Young stars have rotation periods in the range of 0.2 to 10 days \citep{Bouvier2014}, so we randomly choose an initial rotation period $P_0$ from a log-uniform distribution in this range for each age sample. Equation \ref{eqn:gyro} then gives a current rotation period for each sample, which we convert to an equatorial velocity with the stellar radius $R$. We estimate $R$ by interpolating Dartmouth pre main sequence isochrones \citep{Dotter2008} using the companion temperature and system age. We finally convert to projected velocity $v\sin{i}$ by randomly sampling a uniform distribution for the inclination $\sin{i}$.

The gyrochronology relations are invalid for stars with $T \gtrsim 6250\ K$, the canonical limit at which the convective zone is too small to efficiently remove angular momentum to the stellar wind and spin down the star \citep{Pinsonneault2001}. \citet{Zorec2012} fit Maxwellian distributions to the equatorial velocity of A- and B-type stars in several mass bins. For $T > 7000\ K$, we linearly interpolate the fit parameters as a function of mass and sample the resulting Maxwellian probability density function. 

Typical velocities from the gyrochronology relationships are $10-20 \mathrm{km\ s}^{-1}$, while the Maxwellian velocity distributions have typical velocities $\sim 100 \mathrm{km\ s}^{-1}$. We transition between the two regimes for temperatures in the range $6000\ K < T < 7000\ K$ by first estimating the equatorial velocities from the gyrochronology relationship (Equation \ref{eqn:gyro}) at $T=6000\ K$. We then fit the velocities to a Maxwellian distribution, and add the result to the tabulated parameters from \citep{Zorec2012}. With the extended table, we treat stars in the transition range the same way we treat hot stars.
 
\begin{figure}
\includegraphics[width=\columnwidth]{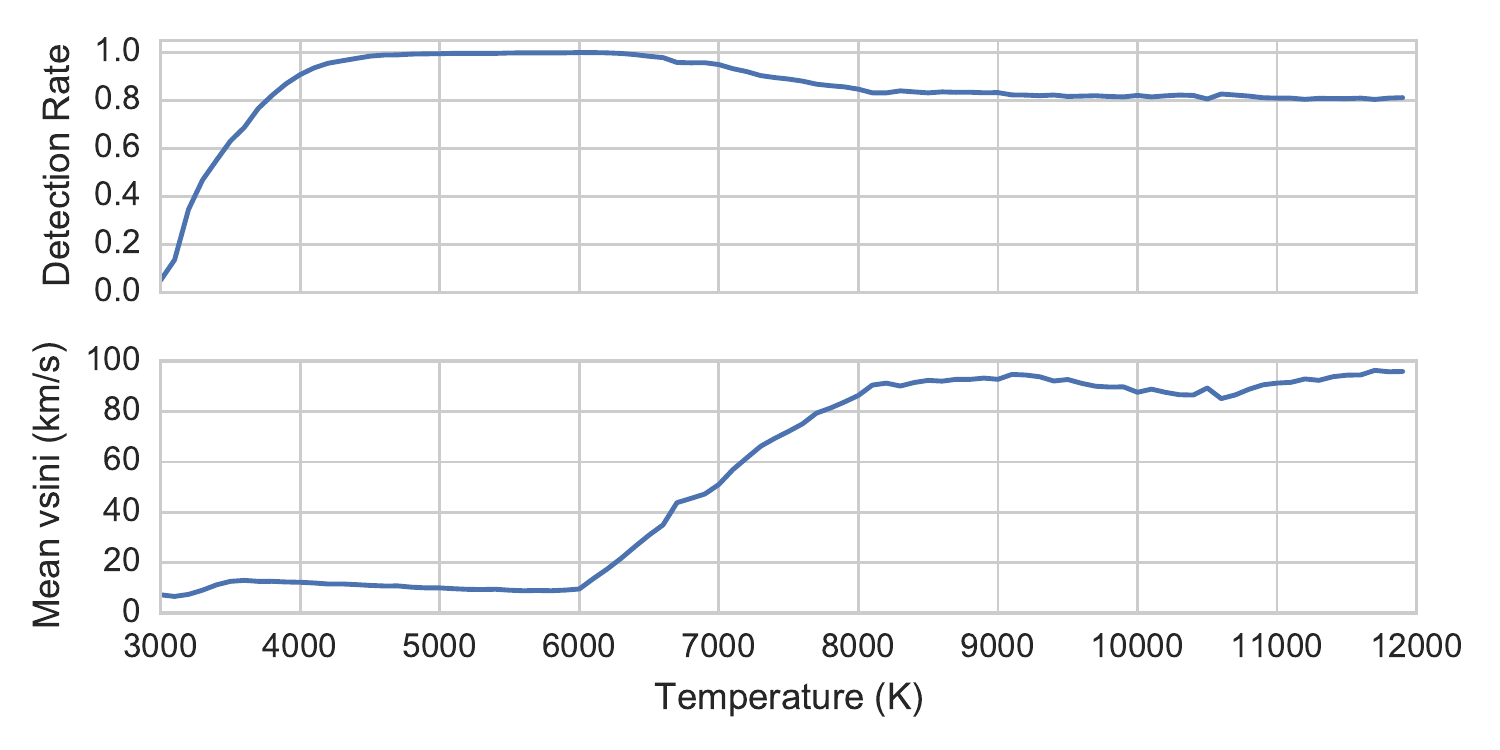}
\caption{Marginalized detection rate for the same star as shown in Figure \ref{fig:detrate_2d}. The fall in detection rate towards hotter stars is caused by the increase in typical rotational speeds.}
\label{fig:marginalized}
\end{figure}

We show the marginalized detection rate and mean value of $v\sin{i}$ as a function of temperature in Figure \ref{fig:marginalized}. Both the detection rate and the average $v\sin{i}$ are smoothly varying, and show the expected behaviour with temperature. The detection rate falls with hotter temperatures because the companions are expected to be fast rotators, which are more difficult to detect.

\subsection{Conversion to Mass Ratio}

The result of the previous analysis is a series of estimates for the detection rate as a function of companion temperature for each observation of each star. We convert companion temperature to mass by interpolating Table 5 of \citet{Pecaut2013}. Next, we estimate the primary mass for each star as the median of the mass samples developed in Section \ref{sec:sp}. We then convert each detection rate curve to be a function of mass ratio ($Q_j(q)$, where $j$ denotes the $j$th star in the sample), and linearly interpolate onto a grid in mass ratio from $0 < q_i < 1$. Finally, we combine the detection rate curves for each star \emph{with no companion detection in our data} into an estimate of the survey-wide completeness by taking the average of the detection rate for all stars:

\begin{equation}
Q(q_i) = \frac{1}{N_i} \sum_j Q_j(q_i)
\label{eqn:completeness}
\end{equation}
In the equation above, $N_i$ is the number of sample stars that contain an estimate for $Q(q_i)$ without extrapolating. For $q_i \sim 0.2$, $N_i$ is near the total sample size. However, $N_i$ falls for both low and high q, since a $3000\ K$/$12000\ K$ companion has a mass ratio $q = 0.08/2.0$ for an A9V primary, but $q = 0.007/0.19$ for a B0V primary. Our sensitivity analysis therefore does not sample large mass ratios around the very early-type primary stars in the sample, and does not sample very low mass ratios around late-type primary stars.

\citet{Gullikson2016} used a very similar method to search for known companions to A- and B-type stars, and found that the detection rate is high for G- and K-dwarf companions but very low for hot companions. The search grid used in this work includes much hotter temperatures, and we have several detections of hot companions (see Table \ref{tab:companions}). We test to determine if the completeness is reasonable at large mass ratios by comparing to known binary systems. We detect 15 of the 25 stars in our sample with a hot ($T > 7000\ K$) companion in either the Washington Double Star Catalog \citep{WDS} or the Ninth Catalog of Spectroscopic Binary Orbits \citep{SB9}. The completeness function for hot, roughly equal-mass companions suggests the probability of detection is $\sim 80\%$, which is still incompatible with our low detection rate. The discrepancy may be due to an underestimate of the typical rotation rates for hot stars, which we use when marginalizing out the dependence on $v\sin{i}$. Additionally, rapidly rotating companions, especially when they have a similar temperature to the primary, are more difficult to detect if they have a small radial velocity offset from the primary star. While the injection and recovery experiments do sample velocity space to account for this, they may be over-sampling companions with very large velocity offsets and producing anomalously high detection rates. We account for the discrepancy by introducing a scaling factor: we multiply the estimated detection rate for all companions with $T > 7000\ K$ by $f=0.8$.

\begin{figure}
\includegraphics[width=\columnwidth]{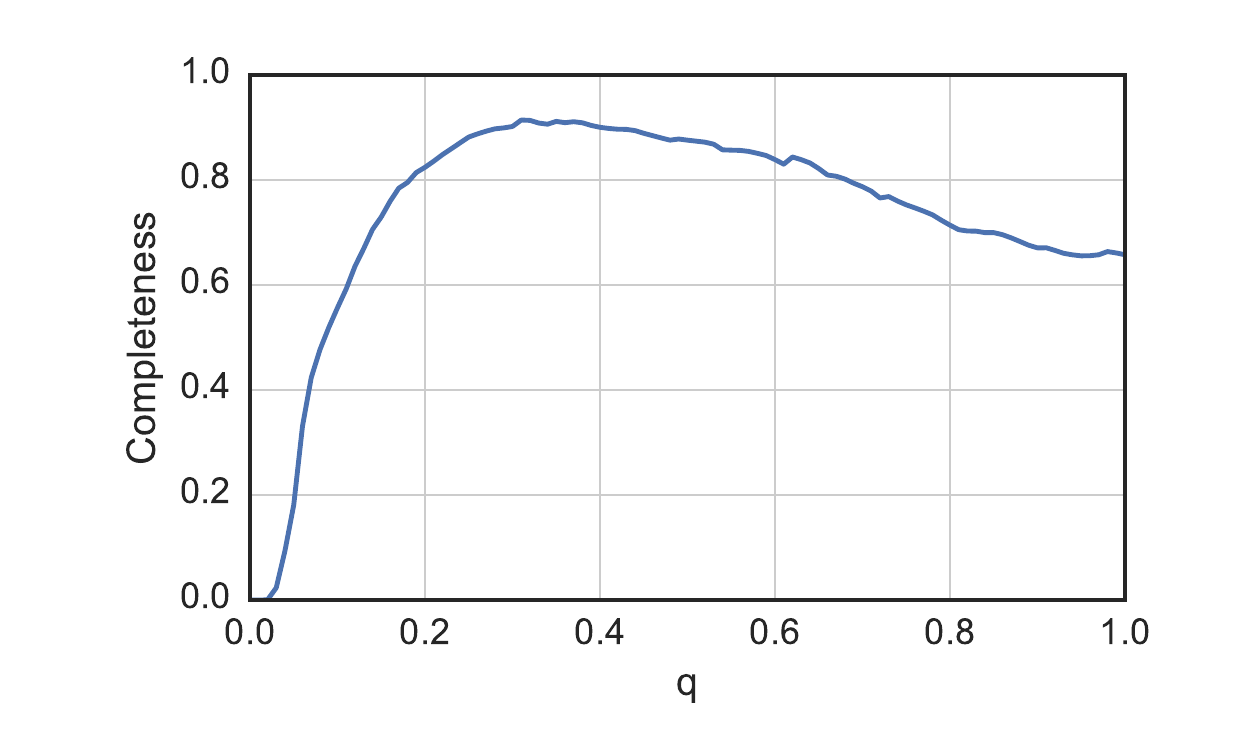}
\caption{Survey completeness as a function of mass ratio ($q$).}
\label{fig:completeness}
\end{figure}

We show the resulting total survey completeness in Figure \ref{fig:completeness}. The completeness falls very rapidly towards low mass ratios, although we are still $\sim 60\%$ complete at $q = 0.1$. The slow fall-off towards large mass ratios is caused by a combination of the scale factor described above and the inherent difficulty of detecting rapidly rotating companions (see Figure \ref{fig:marginalized}). The detection rate at large mass ratios is now $\sim 0.6-0.7$, which is consistent with our 15/25 empirical detection rate. 

\section{Mass-Ratio Distribution}
\label{sec:mrd}

We are now finally in a position to estimate the mass-ratio distribution for our sample. We estimate the mass for each detected companion star by sampling the temperatures given in Table \ref{tab:companions} as a gaussian, and converting each temperature sample into a mass sample. We do the conversion to mass both by interpolating Table 5 from \citet{Pecaut2013}, and by interpolating from temperature and system age (see Section \ref{sec:sp}) to mass with Dartmouth isochrones \citep{Dotter2008}. Both methods give similar results in most cases. Since the isochrone masses are more accurate at young ages, we use them throughout the analysis that follows. We sample the mass ratio of the system by dividing the companion mass samples by samples of the primary mass (Section \ref{sec:sp}). We denote the $n$th mass ratio sample for the $k$th star as $q_k^{(n)}$, and denote the number of these samples as $N_k$.

We do not use systems with more than one companion, unless the wider companion is separated by $ > 10''$ from the primary star. We mark the 50 companions we use in the mass ratio analysis with the fourth column of Table \ref{tab:companions}. Many of the companions we use in the analysis only have one detection in our data; 26/36 of these are previously known companions and so don't need follow-up to confirm. The remaining 10 are new and unconfirmed detections; these all have very strong CCF signals and are likely to be confirmed with follow-up spectroscopy or imaging. Their inclusion does not significantly change the results.

\subsection{Fitting  Methodology}

We use the methodology developed in \citet{Foreman2014} to perform bayesian inference on the shape and form of the mass-ratio distribution. The log-likelihood function in this formalism is derived from modeling the survey as a draw from the inhomogeneous Poisson process with rate density $\Gamma \equiv KQ(q)P(q)$:

\begin{multline}
\ln{\mathcal{L}(\{\vec{x_k}\}| \vec{\theta})} = -K \int_0^1 Q(q)P(q|\vec{\theta})dq + \\ \sum_{k=1}^K \ln{\frac{K}{N_k} \sum_{n=1}^{N_k} Q(q_k^{(n)}) P(q_k^{(n)}|\vec{\theta})}
\label{eqn:money}
\end{multline}

In the above equation, $\{\vec{x_k}\}$ denotes the data for star $k$, and $\vec{\theta}$ denotes the parameters for the model we are fitting. $K=50$ is the number of stars used in the analysis, $Q(q)$ is the completeness function shown in Figure \ref{fig:completeness}, and $P(q|\vec{\theta})$ is the likelihood function for the mass ratio given the model parameters. We fit the data to three distinct distributions: a histogram ($P_1$), a lognormal distribution ($P_2$), and a power law ($P_3$):

\begin{align}
 P_1(q|\vec{\theta}) &= \begin{cases}
      \hfill \theta_1 \hfill & q \in \Delta_1 \\
      \hfill \theta_2 \hfill & q \in \Delta_2 \\
      \hfill \ldots \\
      \hfill \theta_7 \hfill & q \in \Delta_7
     \end{cases} \label{eqn:P1} \\
 %P_2(q|\vec{\theta}) &= \frac{A}{\sqrt{2\pi\sigma^2}} e^{-\frac{(q - \mu)^2}{2\sigma^2}} \label{eqn:P2} \\
 %P_2(q|\vec{\theta}) &= \frac{\Gamma(a+b)}{\Gamma(a)\Gamma(b)} q^{a-1}(1-q)^{b-1} \\
 P_2(q|\vec{\theta}) &= \frac{A}{q\sqrt{2\pi\sigma^2}} e^{-\frac{(\ln{q} - \mu)^2}{2\sigma^2}} \label{eqn:P2} \\
 P_3(q|\vec{\theta}) &= (1-\gamma)q^{-\gamma} \label{eqn:P3}
\end{align}
The constant $A$ in the lognormal distribution is a renormalization factor such that the distribution is only defined from $0 < q < 1$:

\begin{equation}
%A = \frac{2}{\erf\left(\frac{\mu}{\sigma \sqrt{2}}\right) - \erf\left(\frac{\mu - 1}{\sigma \sqrt{2}}\right)}
A = \frac{2}{1-\erf\left(\frac{\mu}{\sigma \sqrt{2}}\right)}
\end{equation}

We fit all distributions via Importance Nested Sampling with the MultiNest code \citep{multinest}. Following \citet{Foreman2014}, we apply a smoothing prior on the parameters $\vec{\theta}$ for the histogram model:

\begin{align}
P(\vec{\theta}| \alpha, m, \tau, \epsilon) &= \mathcal{N}(\vec{\theta} | m, K(\{\Delta_j\}, \alpha, \tau, \epsilon)) \\
K_{ij} &= \sqrt{\left[\alpha \exp{\left(-\frac{(\Delta_i - \Delta_j)^2}{2\tau^2}\right)}\right]^2 + \epsilon^2 \delta_{ij}}
\end{align}
The smoothing prior is an 7-dimensional gaussian with mean $m$ and covariance matrix $K_{ij}$, and encodes our belief that the mass-ratio distribution is a smoothly varying function while leaving enough flexibility to let the data drive the shape of the function. Since we have introduced three new hyperparameters ($a, m, \tau, \epsilon$), we must apply a prior to them and marginalize over them when estimating the bin heights. We choose log-uniform priors for $a, \tau$, and $\epsilon$, and a uniform prior for the mean $m$. The full posterior probability distribution for the histogram model is:

\begin{equation}
P_1(\vec{\theta} | \{\vec{x_k}\}) \propto \mathcal{L}_1(\{\vec{x_k}\}| \vec{\theta}) P(\vec{\theta}| \alpha, m, \tau, \epsilon) P(\alpha, m, \tau, \epsilon)
\end{equation}

The lognormal distribution only has two parameters ($\mu, \sigma$), and was chosen because it has a similar shape to the histogram resulting from the first model. We use uniform priors on both $\mu$ and $\sigma$, although we note that $\mu$ is compared to $\ln{q}$ and so acts like a log-uniform prior. The power law has only one parameter ($\gamma$); we use a uniform prior in the fit.

\subsection{Malmquist Bias Correction}

We are trying to recover the intrinsic distribution from an observed sample, so we must fit the data to the probability distribution function (PDF) for mass ratio, \emph{given that we observed the star}: $P(q|\vec{\theta}, \mathrm{obs})$. In a volume-limited sample, this is equal to $P(q|\vec{\theta})$. However, our sample is magnitude-limited and therefore suffers from Malmquist bias. There is a higher probability for equal-mass binary systems to occur in our survey because they contribute twice the flux and are therefore more likely to fall under the magnitude limit. We can calculate the PDF for mass ratio, given that we observed the system, from Bayes' theorem:

\begin{equation}
P(q|\vec{\theta}, \mathrm{obs}) = \frac{P(\mathrm{obs}|q) P(q|\vec{\theta})}{\int_0^1 P(\mathrm{obs}|q) P(q|\vec{\theta}) dq}
\end{equation}

We already know $P(q|\vec{\theta})$ (Equations \ref{eqn:P1} - \ref{eqn:P3}). We estimate $P(\mathrm{obs}|q)$ by simulating a very large sample of binary stars via these steps:

\begin{enumerate}
\item Draw random primary star masses from the Kroupa IMF \citep{Kroupa2002}
\item Draw a random distance for each star from a disk with infinite extent and scale height of $150$ pc \citep[the approximate scale height of the Milky Way disk for A-type stars,][]{BM1998}. 
\item For each $q$ from 0 to 1, in steps of 0.01:
\begin{enumerate}
  \item Add a companion star to each primary with the appropriate mass to make a binary system with mass ratio $q$.
  \item Calculate the combined absolute V-magnitude by interpolating Table 5 of \citet{Pecaut2013}.
  \item Calculate apparent magnitude $V$ from the absolute magnitude and distance.
  \item Find fraction of stars ($f(q)$) with apparent $V < 6$
\end{enumerate}
\item Fit the sampled fractions $f(q)$ to a 5th-order polynomial.
\end{enumerate}

With the fitted malmquist-correction polynomial, we then substitute $P(q|\vec{\theta}, \mathrm{obs})$ everywhere $P(q|\vec{\theta})$ appears in Equation \ref{eqn:money}.

\begin{figure}
\includegraphics[width=\columnwidth]{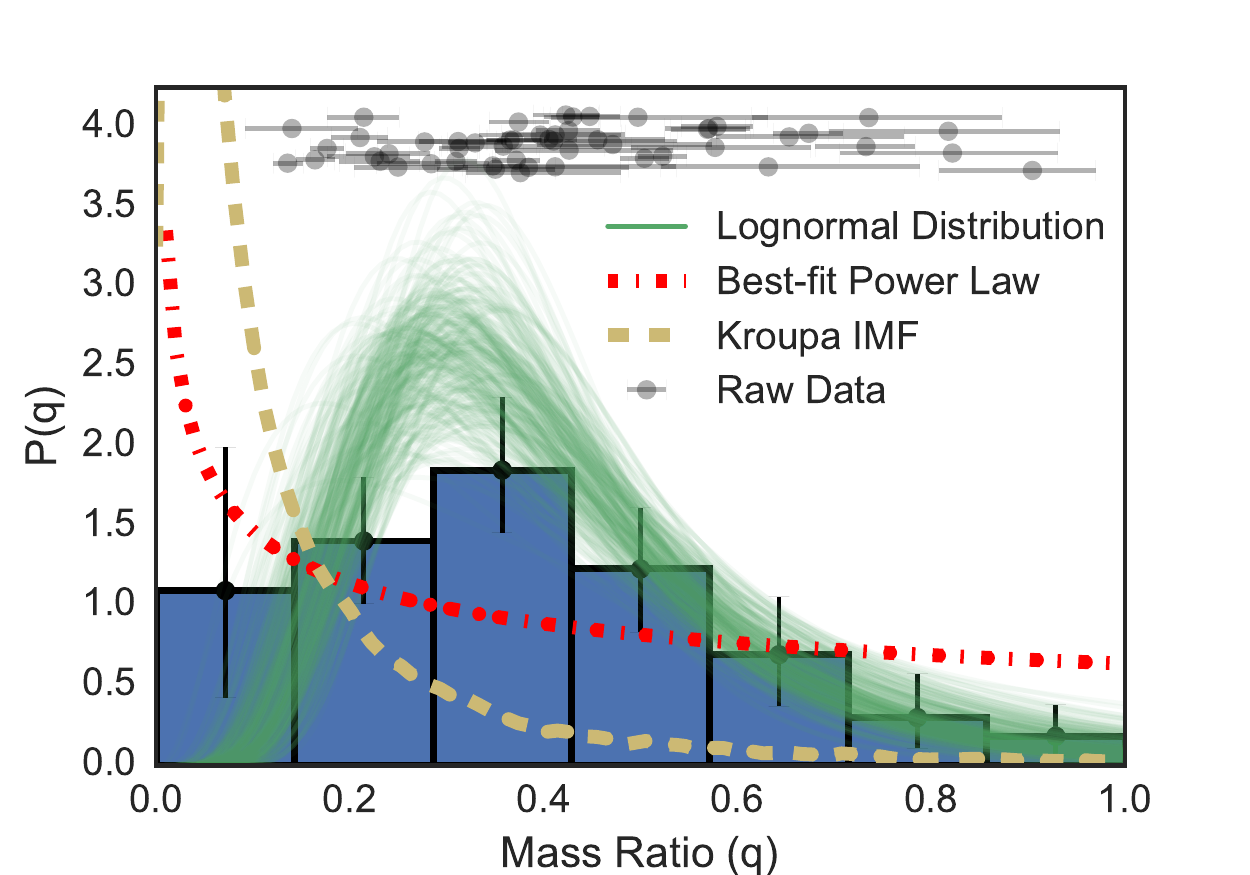}
\caption{Mass-ratio distribution for our sample. The data was fit to a histogram, a lognormal distribution, and a power law. The histogram is shown in the solid blue blocks, with $1 \sigma$ uncertainties marked with error bars. The variance of the lognormal  fit is shown with 300 samples from the posterior probability distribution for the parameters in green. We also show the best-fit power law and the mass-ratio distribution resulting from random pairing of the Kroupa initial mass function. Finally, we display the raw mass ratio measurements with associated uncertainties in the cluster of data points near the top of the figure.}
\label{fig:mrd}
\end{figure}

We summarize the parameters in Table \ref{tab:parameters}, and show the resulting fits in Figure \ref{fig:mrd}. The $1\sigma$ uncertainties in the bin heights from the histogram model are shown as error bars, and we overplot 300 samples of the lognormal distribution fit to show the spread allowed by the data. The best-fit power law is plotted with a red dot-dashed line. We also estimate the mass-ratio distribution expected from random pairing of the Kroupa Initial Mass Function (IMF), and show the result in yellow. We estimate the distribution by drawing 100000 primary stars from the IMF with masses between $1.5 < M/M_{\odot} < 20$. We then draw companions from the same IMF, with the restriction that the companion has a lower mass than the primary. The result plotted in yellow in Figure \ref{fig:mrd} is a gaussian kernel density estimate of the resulting mass ratios, with a bandwidth of $0.05$.

\section{Discussion}
\label{sec:discussion}

Previous measurements of the mass-ratio distribution find that the data is well fit by a power law. \citet{Kouwenhoven2007} compiled spectroscopic, imaging, and astrometric observations of binary stars with intermediate-mass primaries in the Scorpius OB association, and derived a power law index of  $-0.45 \pm 0.15$. More recently \citet{DeRosa2014} performed an adaptive optics and common proper motion search for companions to field A-type stars. They found that the distribution for companions on wide ($a > 125\ AU$) orbits has a very steep power law index of $-2.3^{+1.0}_{-0.9}$, while the distribution for close ($30\ AU < a < 125\ AU$) companions is consistent with flat.

\subsection{Model Comparison}

The most striking feature of the mass-ratio distribution shown in Figure \ref{fig:mrd} is the turnover or flattening at intermediate q. The maximum of the lognormal distribution occurs at $q = 0.30 \pm 0.03$ and is an estimate of the characteristic scale.

Although the power law fit and the Kroupa IMF are visually very poor fits to the data, we formally compare the models to ensure that the different form is not an artifact of binning or simple noise. We make the comparison using the posterior odds:

\begin{equation}
\mathrm{Odds} = \frac{\int_{\vec{\theta_1}} P_1(q|\vec{\theta_1}) P(\vec{\theta_1})}{\int_{\vec{\theta_2}} P_2(q|\vec{\theta_2}  P(\vec{\theta_2})} \equiv \frac{Z_1}{Z_2}
\end{equation}

The integrals are estimated as part of the nested sampling algorithm in the MultiNest code. The odds ratio comparing the lognormal distribution to power law fit is $Z_\mathrm{lognormal} / Z_\mathrm{power} = 5.1 \pm 0.1 \e{6}$, indicating a very strong preference for the lognormal distribution model. We also compare to the mass-ratio distribution expected for random pairing from the Kroupa IMF and to a uniform distribution (a special case of the power law). In these cases, there are no free parameters so the evidence integral just becomes the likelihood function (Equation \ref{eqn:money}). The corresponding odds ratios are  $Z_\mathrm{lognormal} / Z_\mathrm{IMF} = 6.5 \pm 0.1 \e{22}$ and $Z_\mathrm{lognormal} / Z_\mathrm{uniform} = 7.0 \pm 0.1 \e{6}$. Both of these again demonstrate a very strong preference for the lognormal distribution. 

The extreme unlikeliness of the Kroupa IMF model also indicates that our sample is not significantly biased by foreground or background contaminants. In fact, the present-day background star mass function is more bottom heavy than the initial mass function because some of the massive stars have evolved to white dwarfs or ended their lives in a supernova. The comparison to a Kroupa IMF therefore \emph{underestimates} the likelihood of background star contamination.

\subsection{Comparison to Previous Results}

Our mass-ratio distribution appears to be in tension with the results of the VAST survey \citep{DeRosa2014}, which finds a nearly flat distribution for close companions. However, their subsample of close companions only includes 18 binaries, so it is possible that the different forms are just a result of small number statistics. To assess the degree of tension, we use the Anderson-Darling  test \citep{Anderson1954} to find the probability that both their close companion subsample and our companions are drawn from the same parent distribution. We only use companions from this work with mean $\overline{q} > 0.15$ because the VAST survey subsample makes the same cut. The VAST survey also only included stars with projected separations $a < 125$ AU in their close companion subsample. Since we cannot estimate the separation from our data, we do not make such a cut. We could make a cut using the \emph{maximum} possible separation, set by the distance and spectrograph slit width, but doing so vastly reduces the number of detections and does not affect the result. 

To account for measurement uncertainties in the mass ratios, we draw from both our mass ratio samples ($q_k^{(n)}$, see Section \ref{sec:mrd}) and the VAST mass ratio values many times and compute the Anderson-Darling test statistic each time. Since \citet{DeRosa2014} do not quote uncertainties, we assume uncertainties of $\sigma_q = 0.05$ for all of their measurements. The result is $p = 0.10^{+0.07}_{-0.04}$; we cannot reject the hypothesis that both samples come from the same distribution.

\subsection{Theoretical Implications}
\label{subsec:theory}

The mass-ratio distribution derived in this work has a very different form than the power law found for companions at wide separations. This is likely a result of disk interactions as the two components are accreting. The close companions that we detect may form with similar masses to their counterparts at large separations ($a \gtrsim 1000\ AU$), but preferentially accrete matter from the dense primary star disk. The result would be a depletion of low mass ratio companions as they become intermediate to high mass ratio companions. The characteristic scale of $\sim 0.3$ that we see in Figure \ref{fig:mrd} would then be related to the disk timescale, since with enough time the preferential accretion would push all companions to $q = 1.0$. 

It is also possible that some of the companions found in this work were formed from a gravitationally unstable disk \citep[e.g.][]{Kratter2006, Stamatellos2011}. Being a completely different formation mechanism than the way wide companions form, we would expect the initial companion mass function to differ. Such companions would undergo the same preferential accretion discussed above. 

Large scale simulations are likely needed to distinguish between the two scenarios and fully interpret the results of this survey. A significant amount of work has already been put towards this end in the form of radiation hydrodynamic simulations of giant molecular clouds \citep{Bate2012, Krumholz2012}. However, the present simulations do not generate enough stars more massive than the sun to quantitatively compare binary and multiple star statistics to observations.

\section{Summary}

In this work, we described a binary survey of 341 bright A- and B-type stars. We used the direct spectral detection method \citep{Gullikson2016} to find the spectral lines of 64 companions with temperatures ranging from $3600 - 16000\ K$.  We used the cross-correlation functions to estimate the temperature and surface gravity of most of our sample stars, and converted to mass and age by interpolating Padova stellar evolutionary tracks \citep{Bressan2012}. Likewise, we convert the companion temperature measurements to mass by using solar metallicity Dartmough evolutionary tracks \citep{Dotter2008}. 

We then use the formalism introduced in \citet{Foreman2014}, which self-consistently accounts for measurement errors, to infer the form of the mass-ratio distribution (shown in Figure \ref{fig:mrd}). Unlike most previous work, we find that a power law is a poor descriptor of the data and find that a lognormal distribution performs much better. This result, which only includes close companions since it is a spectroscopic technique, is consistent with the 18 close companions found in the VAST survey \citep{DeRosa2014}. However, this result shows much more detail due to a larger number of companions. 

We interpret the mass-ratio distribution in terms of formation mechanism in Section \ref{subsec:theory}. It is likely that the mass-ratio distribution we find is largely a result of preferential accretion onto the secondary star, which largely stops when the circumprimary or circumbinary disk dissipates.

In the effort of open and reproducible research, we have made several data products freely available to the community. All of the reduced and telluric-corrected spectra used in this study are available at \url{https://zenodo.org/record/46340}. Samples of the primary and companion mass and system age posterior distributions are available at \url{https://zenodo.org/record/48073}, as are the posterior distributions for the parameters fit in Section \ref{sec:mrd} and every cross-correlation function generated in our analysis. We additionally provide a series of python libraries and jupyter notebooks with the computer code we used for the analysis on github: \url{https://github.com/kgullikson88/BinaryInference}.

\section*{Acknowledgements}
We would like to thank the anonymous referee for their helpful comments, which have notably improved this work. This research has made use of the SIMBAD database, operated at CDS, Strasbourg, France, and of Astropy, a community-developed core Python package for Astronomy (Astropy Collaboration, 2013).
It was supported by a start-up grant to Adam Kraus as well as a University of Texas Continuing Fellowship and a Dissertation Writing Fellowship to Kevin Gullikson.

This work used the Immersion Grating Infrared Spectrograph (IGRINS) that was developed under a collaboration between the University of Texas at Austin and the Korea Astronomy and Space Science Institute (KASI) with the financial support of the US National Science Foundation under grant AST-1229522, of the University of Texas at Austin, and of the Korean GMT Project of KASI.

The Hobby-Eberly Telescope (HET) is a joint project of the University of Texas at Austin, the Pennsylvania State University, Stanford University, Ludwig-Maximilians-Universit\"at M\"unchen, and Georg-August-Universit\"at G\"ottingen. The HET is named in honor of its principal benefactors, William P. Hobby and Robert E. Eberly.

Based on observations at Cerro Tololo Inter-American Observatory, National Optical Astronomy Observatory (NOAO Prop. IDs: 13A-0139, 13B-0112, 2014A-0260, 14A-0260, 15A-0245; PI: Kevin Gullikson), which is operated by the Association of Universities for Research in Astronomy (AURA) under a cooperative agreement with the National Science Foundation. 

\clearpage
\newpage

\LongTables
\begin{deluxetable*}{|l|lrrrrllllll|}
\tabletypesize{\scriptsize}
\tablewidth{0pt}
\tablecaption{Sample Properties}
\tablehead{
      \colhead{} & \colhead{} & \colhead{} & \colhead{} &  \colhead{}  & \colhead{parallax} & \colhead{$T_\mathrm{eff}$} & \colhead{$\log{g}$} & \colhead{Mass} & \colhead{Age} & \colhead{} \\
       \colhead{star} & \colhead{SpT} & \colhead{RA} & \colhead{DEC} &  \colhead{V} &   \colhead{
       (mas)} & \colhead{(K)} & \colhead{(cgs)} & \colhead{($M_{\odot}$)} & \colhead{(Myr)} & \colhead{Ref}
       }

\startdata

     HIP 813 &     B9Vn &    00:10:02.20 &   +11:08:44.93 &   5.537 &     10.68 &   $12516 \pm 426$ &  $4.3 \pm 0.14$ &  $3.1^{+0.18}_{-0.17}$ &      $85^{+56}_{-52}$ &       1 \\
    HIP 1191 &    B8.5V &    00:14:54.52 &   -09:34:10.45 &   5.757 &      9.63 &  $12000 \pm 1000$ &  $4.5 \pm 0.25$ &  $2.8^{+0.41}_{-0.37}$ &      $23^{+63}_{-16}$ &       2 \\
    HIP 1366 &      A2V &    00:17:05.50 &   +38:40:53.89 &   4.610 &     10.56 &    $9371 \pm 319$ &  $4.0 \pm 0.14$ &  $2.2^{+0.18}_{-0.16}$ &    $464^{+83}_{-119}$ &       1 \\
    HIP 1647 &      B9V &    00:20:39.04 &   -69:37:29.68 &   5.498 &     10.25 &   $11393 \pm 387$ &  $4.0 \pm 0.14$ &  $2.8^{+0.21}_{-0.18}$ &     $206^{+48}_{-79}$ &       1 \\
    HIP 2381 &      A3V &    00:30:22.65 &   -23:47:15.65 &   5.190 &     18.83 &    $8364 \pm 284$ &  $4.0 \pm 0.14$ &  $1.9^{+0.16}_{-0.13}$ &   $715^{+135}_{-183}$ &       1 \\
    HIP 2505 &     B8Vn &    00:31:46.36 &   +54:31:20.23 &   4.732 &      8.64 &  $12000 \pm 1000$ &  $4.0 \pm 0.25$ &  $2.9^{+0.45}_{-0.40}$ &     $58^{+104}_{-48}$ &       2 \\
    HIP 2548 &    B9.5V &    00:32:23.78 &   +06:57:19.66 &   5.698 &     12.35 &   $11864 \pm 403$ &  $4.4 \pm 0.14$ &  $2.8^{+0.16}_{-0.15}$ &      $77^{+69}_{-51}$ &       1 \\
    HIP 3300 &      B2V &    00:42:03.90 &   +50:30:45.09 &   4.810 &      2.28 &  $18000 \pm 1000$ &  $4.0 \pm 0.25$ &  $5.7^{+0.67}_{-0.63}$ &      $19^{+19}_{-13}$ &       2 \\
\enddata
\tablecomments{The spectral types, coordinates, V-band magnitudes, and parallax measurements are taken from the Simbad database; the spectral type given is that of the brightest star if part of a known multiple system. The "Ref" column denotes the reference for the stellar effective temperature, surface gravity, mass, and age. The references are: [1]: \citep{David2015}; [2]: This study.}
\label{tab:sample}
\end{deluxetable*}

\clearpage
%\newpage     

%\LongTables
\begin{deluxetable}{|l|rlrr|}
\tabletypesize{\small}
\tablewidth{0pt}
\tablecaption{Spectroscopic Observation Log}
\tablehead{
      \colhead{} &  \colhead{MJD} & \colhead{}  & \colhead{exptime} & \colhead{} \\
       \colhead{star} &  \colhead{(JD-2450000)} &   \colhead{instrument} & \colhead{(sec)} & \colhead{snr}
       }
       
\startdata
     HIP 813 &  6618.58 &       TS23 &      557 &  1810 \\
    HIP 1191 &  6552.60 &     CHIRON &     1200 &   320 \\
    HIP 1366 &  6585.73 &       TS23 &      274 &  1790 \\
    HIP 1647 &  6510.74 &     CHIRON &     1105 &   630 \\
    HIP 2381 &  6874.71 &     CHIRON &      505 &   580 \\
    HIP 2505 &  6586.71 &       TS23 &      235 &  1700 \\
    HIP 2505 &  6962.71 &       TS23 &      804 &  1960 \\
    HIP 2505 &  6945.71 &     IGRINS &      224 &   210 \\
\enddata
\tablecomments{See Section \ref{subsec:specdata} for details of the signal-to-noise ratio calculation (the snr column)}
\label{tab:observations}
\end{deluxetable}

\clearpage

\begin{deluxetable*}{|l|llllllll|}
\tabletypesize{\small}
\tablewidth{0pt}
\tablecaption{NIRI Observation log}
\tablehead{
       \colhead{star} &  \colhead{K} &   \colhead{date} & \colhead{Exposure} & \colhead{$\rho$} & \colhead{$\theta$} & \colhead{$\Delta \mathrm{K_c}$} & \colhead{$M_2 (M_{\odot})$} & \colhead{$a$ (AU)} 
       }
       
\startdata
  HIP 22833 &  4.83 &  2015-03-29 &   50x0.3 & $0.352 +/- 0.003$ & $14.1 \pm 0.7$ & $3.04 \pm 0.06$ & $0.9^{+0.16}_{-0.17}$ & $25.1 \pm 0.24$ \\
  HIP 22958 &  5.79 &  2015-10-19 &   52x0.4 & $0.445 +/- 0.003$ & $65.0 \pm 0.3$ & $2.57 \pm 0.05$ & $1.5^{+0.35}_{-0.28}$ & $124.1 \pm 0.81$ \\
  HIP 26093 &  5.96 &  2015-03-23 &   33x0.6 & \nodata & \nodata & \nodata & \nodata & \nodata \\
  HIP 26126 &  5.07 &  2015-03-29 &   40x0.4 & \nodata & \nodata & \nodata & \nodata & \nodata \\
  HIP 29735 &  5.10 &  2015-03-22 &   83x0.2 & \nodata & \nodata & \nodata & \nodata & \nodata \\
  HIP 31278 &  5.46 &  2015-03-28 &   29x0.6 & \nodata & \nodata & \nodata & \nodata & \nodata \\
  HIP 38538 &  4.66 &  2015-10-19 &   75x0.2 & \nodata & \nodata & \nodata & \nodata & \nodata \\
  HIP 39847 &  4.66 &  2015-03-27 &   50x0.3 & \nodata & \nodata & \nodata & \nodata & \nodata \\
\enddata
\tablecomments{$\dagger$: There is no K-magnitude tabulated in the Simbad Database. The value quoted here is estimated from the spectral type of the star and its V-band magnitude.}
\label{tab:imaging_obs}
\end{deluxetable*}

\clearpage
\newpage

\LongTables
\begin{deluxetable*}{|l|rrlrrrll|}
\tabletypesize{\small}
\tablewidth{0pt}
\tablecaption{Companion Detections}
\tablehead{
     \colhead{} & \colhead{} & \colhead{} & \colhead{Used in} & \colhead{$T_\mathrm{eff}$} & \colhead{$v\sin{i}$} & \colhead{[Fe/H]} & \multicolumn{2}{c}{Mass ($M_{\odot}$)} \\ 
       \colhead{star} & \colhead{component} & \colhead{$N_\mathrm{obs}$} & \colhead{Analysis}  &   \colhead{(K)} &   \colhead{($\mathrm{km\ s}^{-1}$)} &  \colhead{(dex)} &   \colhead{Isochrone} &   \colhead{Spectral Type}
       }
       
\startdata
   HIP 2548 & Aa,Ab &     2 & yes &  $5732 \pm 112$ &      30 &    -0.5  &  $1.0^{+0.06}_{-0.04}$ &  $1.0^{+0.04}_{-0.03}$ \\
   HIP 5348 & AB &     1 & no &  $12000 \pm 1000$ &     75 &     0.0 &  $2.9^{+0.40}_{-0.36}$ &  $3.3^{+0.23}_{-0.29}$ \\
     HR 545 & Aa,Ab$^{\dagger}$ &     3 & yes &   $4312 \pm 87$ &       5 &    -0.5  &  $0.8^{+0.11}_{-0.10}$ &  $0.7^{+0.01}_{-0.02}$ \\
     HR 604 & AB &     1 & no &  $4736 \pm 150$ &       5 &     0.5  &    $0.9^{+0.17}_{-0.12}$ &  $0.8^{+0.02}_{-0.02}$ \\
  HIP 10732 & AB$^{\dagger}$ &     2 & yes &  $5578 \pm 109$ &       5 &     0.0  &  $1.0^{+0.29}_{-0.04}$ &  $1.0^{+0.02}_{-0.04}$ \\
  HIP 12332 & AB$^{\dagger}$ &     2 & yes &  $5551 \pm 107$ &      10 &     0.0  &    $1.0^{+0.03}_{-0.03}$ &  $1.0^{+0.01}_{-0.04}$ \\
  HIP 12706 & AB &     2 & yes &  $6051 \pm 107$ &      10 &     0.0  &  $1.1^{+0.03}_{-0.03}$ &  $1.1^{+0.04}_{-0.05}$ \\
  HIP 13165 & Aa,Ab &     1 & no &  $5773 \pm 163$ &       5 &    -0.5  &  $1.1^{+0.31}_{-0.08}$ &  $1.0^{+0.10}_{-0.03}$ \\
\enddata
 
 \tablecomments{$\dagger$: Newly discovered companions.}
 \label{tab:companions}
\end{deluxetable*}

\clearpage
\newpage

\LongTables
\begin{deluxetable}{|r|rlll}
\tabletypesize{\normalsize}
\tablewidth{0pt}
\tablecaption{Mass-Ratio Distribution Parameters.}
\tablehead{
     \colhead{} & \colhead{} & \colhead{ } & \multicolumn{2}{c}{$1 \sigma$ credible interval} \\
     \colhead{Model} & \colhead{Parameter} & \colhead{Median Value} & \colhead{Lower} & \colhead{Upper}
       }
       
\startdata

Histogram   & $\theta_1$                    & 0.50 & 0.18 & 1.00 \\
                   & $\theta_2$                    & 1.53 & 1.14 & 1.89 \\
                   & $\theta_3$                    & 2.17 & 1.80 & 2.65\\
                   & $\theta_4$                    & 1.42 & 0.98 & 1.81 \\
                   & $\theta_5$                    & 0.74 & 0.40 & 1.12 \\
                   & $\theta_6$                    & 0.29 & 0.10 & 0.55 \\
                   & $\theta_7$                    & 0.16 & 0.06 & 0.33 \\ \hline
Lognormal  & $\mu$                           & -0.99 & -1.06 & -0.92 \\
                   & $\sigma$                      & 0.46  &  0.40  &  0.54 \\
                   & maximum$^{\dagger}$ & 0.30  & 0.27   & 0.33 \\ \hline
Power Law & $\gamma$                    & 0.29  & 0.15   & 0.42
 
 \enddata
 
 \tablecomments{$\dagger$: This parameter is derived from the mean and standard deviation, not fit.}
 \label{tab:parameters}
\end{deluxetable}

\clearpage
\newpage

\newpage
\clearpage
\bibliography{references}

\end{document}